\documentclass{aa}
\usepackage{natbib}
\usepackage{amsmath}
\usepackage{txfonts}
\usepackage{graphicx}
\def\teff{T_{\rm eff}}
\def\mast{M_\ast}

\def\msol{\mathrm{M}_\odot}
\def\lsol{\mathrm{L}_\odot}

\def\last{L_\ast}

\def\llsol{L/\mathrm{L}_\odot}

\def\cv{c_V}

\def\Prad{P_{\mathrm{rad}}}

\def\g{{\it g\/}}

\def\diff{{\mathrm d}}

\def\unity{ \hbox{1\kern-.23em l} }
\def\zero{ \hbox{0\kern-.23em |} }
\def\field{ \hbox{I\kern-.23em K} }
\begin{document}

%%%%%%%%%%%%%%%%%%%%%%%%%%%%%%%%%%%%%%%%%%%%%%%%%%%%%%%%%%%%%%%%%%%%%%%%%%%%%%
   \title{On the excitation of PG1159-type pulsations}
   \author{     A. Gautschy\inst{1} 
           \and L.G. Althaus\inst{2} 
           \and H. Saio\inst{3} }
   \institute{
         4410 Liestal, Wetterchr\"uzstr. 8c, Switzerland
         %% ETH-Library, R\"amistrasse 101, 8092 Z\"urich, Switzerland
         \and
         Departament de F\'{\i}sica Aplicada, 
         Universitat Polit\`ecnica de Catalunya, Av. del Canal
         Ol\'{\i}mpic, s/n, $08860$, Castelldefels, Barcelona, Spain
         \and
         Astronomical Institute, School of Sciences, Tohoku University, 
         Sendai $980-8578$, Japan
        }
   \offprints{A.~Gautschy, \email{alfred@gautschy.ch}}
   \date{Received 6.XII.2004 / Accepted 17.IV.2005}
%__________________________________________________________________________
%
\abstract{ Stability properties are presented of dipole 
           and quadrupole nonradial oscillation modes of model stars
           that experienced a late helium shell flash on their way to
           the white-dwarf cooling domain. The computed instability
           domains are compared with the observed hot variable central
           stars of planetary nebulae and the GW Vir pulsators.
          }
\maketitle
%_________________________________________________________________________
%-----------------------------------
\section{Introduction}
\label{sect:intro}
%-----------------------------------
The family of pre-white dwarfs (PWDs) referred to as PG$1159$ stars
is spectroscopically defined by the dominance of He~II, C~IV, and O~VI
(and sometimes N~V) lines and a strong H deficiency. Surface mass
abundances of about $33\%$~He, $50\%$~C and $17\%$~O are reported as
typical.  Currently, 32 stars are members of the PG$1159$ family, the
effective temperatures range from about $170\,000$ to
$80\,000$~K. Roughly half of the PG$1159$ stars are embedded in a
planetary nebula.

%.........................................................................
%\endnote{Spectroscopy still allows for up to $5 \%$ H in mass in the
%atmospheres of PG$1159$ stars. Below this threshold, stars are
%considered as H free (Dreizler, Baltic Astronomy, 7, 71, 1998).}
%.........................................................................
Until recently, a hydrogen-deficient composition of the PG$1159$ kind
could not be reconciled with stellar evolution scenarios. Only the
introduction of appropriate overshooting and mixing procedures in
modeling post-AGB stars undergoing a very late thermal pulse (VLTP)
led to PWDs with H-deficient surface layers that are in agreement with
PG$1159$ spectroscopy \citep[][and references therein]{herwig01}.

Ten of the PG$1159$ stars are pulsating variables (also baptized as GW
Vir variables, after the variable-star designation of the prototype
PG$1159$-035; we will use both names synonymously throughout the
text). Four of the ten variable stars have high surface gravities
($\log g > 7$) and low luminosities, i.e. they are essentially on the
corresponding white-dwarf cooling track and have hence evolved around
the ``evolutionary knee'' (cf. Fig.~\ref{fig:hrdm27}).
%......................................................................
%\endnote{The high-$\log g$ variables are also referred to 
%as DOV, i.e. as variable DO stars. This designation, although
%historical, is inappropriate as spectroscopically the stars are
%\emph{not} DO stars.}.
%......................................................................
The rest of the GW Vir stars has low $\log g$ (i.e. $\log g < 7$), in
four cases associated planetary nebulae are known.
%......................................................................
%\endnote{The pulsating planetary nebulae nuclei 
%among the GW Vir stars are frequently called PNNV; calling them lgEV,
%i.e. variable stars among the spectroscopic lgE class (K. Werner,
%1992, in The Atmospheres of Early-Type Stars, p.~278, LNP 401,
%Eds. U.~Heber and C.S.~Jeffery, Springer, 1992) would be much more
%appropriate. The use of PNNV is misleading since we know PNNVs that
%are not PG$1159$ stars. Furthermore, using the designation lgEV would
%also incorporate those GW Vir variables with low~$\log g$ but without
%a PN (such as HS2324 and HE1429). This is all phenomenology, and we
%should not spend too much energy on dreaming up fancy names.
%Astro\emph{physically} we do not need to distinguish between the
%low-gravity PG$1159$ variables and the high-gravity brothers. The
%excitation physics is the same in both cases.}
%...................................................................... 
The observed periods range from about $5$ to some $17$ minutes on the
high-$\log g$ branch and they can exceed one hour in the low-$\log
g$ PG$1159$ variables. Based on these time-scales, the pulsations must
be attributed to intermediate to high-order \g~modes.

The excitation physics of the PG$1159$-type oscillations is still a
matter of debate; this, despite the relative simplicity of the
microphysics in the destabilization region as compared with
e.g. classical pulsating variables. Most of all, in the case of the
PWD pulsations, the stability analyses are not discredited by the
insufficiently understood r{\^o}le of convection which is usually
present when the $\kappa$-mechanism is at work. Nonetheless, no
general agreement exists regarding in particular the chemical
composition in the driving zone
\citep[e.g.][]{starrfieldetal83,SaioPG96,bradleydziem96,GaPG97,anc03,
quirionetal04}.

\citet{SaioPG96} and \citet{GaPG97} studied qualitatively the influence of
the new generation of opacity data, specifically of the OPAL
opacities, on the excitation of GW Vir pulsations. Due to the lack of
evolutionary models with realistic histories that led them through the
thermally pulsing AGB phase, the authors performed their
\emph{explorative computations} on model stars that evolved off the helium
main sequence. The evolutionary phases before the knee differed
significantly from post-AGB tracks. The post-knee epochs
though, converged surprisingly well towards the loci of full
evolutionary computations. The PG$1159$-type instability regions deduced
from the simplified models agreed decently well with observations. The
computed period separations, however, differed sufficiently from the
observed ones to arouse considerable criticism
\citep{obrien00,anc03}. However, it seems to have largely gone 
unrecognized that usually the \emph{excitation} physics is much more
robust than the asteroseismic signatures and that the studies of
\citet{SaioPG96} and \citet{GaPG97} served a useful purpose
nonetheless.

In the following paper we return to the excitation physics of PG$1159$
stars. We set out to show that the Saio and Gautschy results are
robust, as it has also been shown recently in an independent approach
by \citet{quirionetal04}. This time, we go beyond the use of
simplified evolutionary models or the use of envelope models to
describe GW~Vir pulsations. Numerical stability analyses were
performed on state-of-the-art model stars that evolved through a late
helium shell flash after they evolved through the thermally pulsing
AGB, starting originally from the main sequence. Hence, this paper
closes the credibility gap left open by the earlier work.
Section~\ref{sect:moto} introduces the stellar models and reiterates
on the computational tools used for the stability analyses.
Section~\ref{sect:results} presents the results of the stability
computations for dipole and quadrupole modes; in particular the
instability domains are described. The results are discussed and
compared with earlier work in Sect.~\ref{sect:discuss}; the
conclusions in Sect.~\ref{sect:conclude} point out further steps that
need to be taken.
\medskip

\section{Models and Tools}
\label{sect:moto}
%-----------------------------------
%---------------------------------------------------------------------------
%  side-captioned plot
\begin{figure}
      \resizebox{\hsize}{!}{\includegraphics{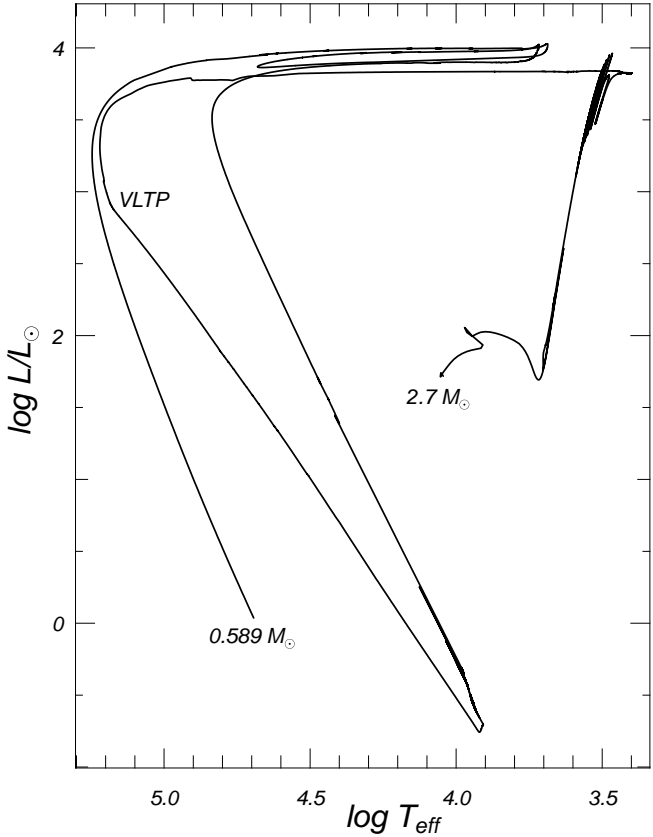}}
      \caption{Evolutionary track of the initially $2.7 \msol$ star
      passing through the AGB, a very late thermal pulse (VLTP), and entering
      the final cooling phase as a hydrogen-deficient white dwarf with
      $0.59 \msol$. Pulsation computations were performed at late
      epochs along the locus tracing the terminal departure from the
      red-giant domain.
%      The evolutionary track from the main
%      sequence to the end of the thermally-pulsing AGB phase occupies
%      only a small region on the cool side of the HR plane as compared
%      with the VLTP-induced born-again red-giant excursion. Only after
%      an additional loop at essentially constant high luminosity does
%      the now hydrogen-free star embark on its terminal voyage towards
%      the white dwarf cooling track.  
}
\label{fig:hrdm27}
\end{figure}
%---------------------------------------------------------------------------

\begin{figure}
      \resizebox{\hsize}{!}{\includegraphics{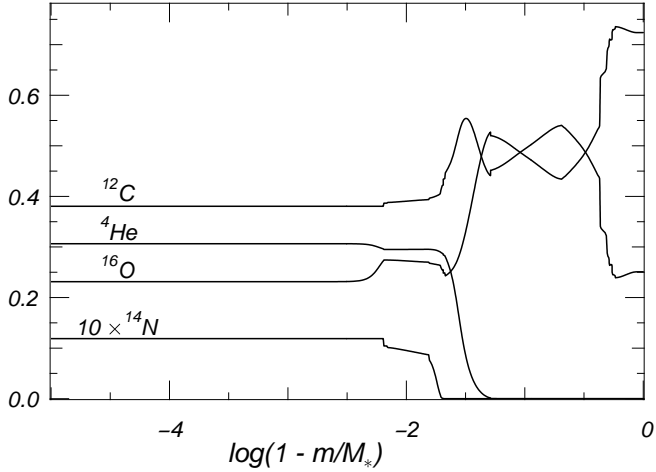}}
      \caption{Representative spatial abundance profiles for selected
      nuclear species for a $0.59 \msol$ model at $\log \teff = 5.15$
      and $\log \llsol = 2.377$. The outermost $10^{-5} M_\ast$ are
      not shown as the compositions do not change there anymore; this
      applies to the evolutionary phases prior to $\log\teff = 5.0$
      along the white dwarf cooling track. Overshoot episodes during
      core helium burning and the AGB evolution left clear signatures
      in the abundance profiles: the step at $-0.3$ and the
      mountain-like feature at around $-1$ of the abscissa.}
\label{fig:chipg1159}
\end{figure}

The model stars on which stability analyses were performed were
extracted from the computations described in
\citet{althausetalvltp05}; they followed an initially $2.7 \msol$ 
star with solar composition from the zero-age main sequence, through
the AGB and a VLTP, to the final cooling sequence as a
hydrogen-deficient white dwarf. Except for the final white-dwarf
cooling phase, the admixture of heavy elements in the star's envelope
was assumed to be solar; the OPAL opacity tables that entered the
computations are described in \citet{althausetalvltp05}. The
abundance evolution of $16$ chemical elements was included via a
time-dependent numerical scheme that simultaneously treated nuclear
evolution and mixing processes due to convection and
overshooting. Such a treatment is particularly important during the
short-lived phase of the VLTP and the born-again episode for which the
assumption of instantaneous mixing is inadequate. In the course of the
VLTP phase, most of the residual hydrogen envelope is engulfed by the
deep helium-flash induced convection zone and is completely
burned. The star is then forced to evolve rapidly back to the
red-giant region and eventually into the domain of the central stars
of planetary nebulae at high effective temperatures but this time as a
hydrogen-deficient, quiescent helium-burning object. We note that the
inclusion of overshooting below the helium-flash convection zone
leads naturally to surface abundances in agreement with those observed
in PG$1159$ stars (see Fig.~\ref{fig:chipg1159}). In particular, also
the N abundance is compatible with the range of observed ones
%.......................................................................
%\endnote{\citet{dreizlerheber98} give number ratios N/He = 0.01 for
%the four GW Vir stars: PG$1159$, PG2131, PG1707 and PG0122. On the other
%hand, for the non-pulsating PG$1159$ stars N/He $<10^{-4}$.
%\citet{werner01} mentions PG$1159$ with a mass
%abundance of N of $0.01$ in his Table~1}
%.......................................................................
(of the order of 0.01 in mass) in pulsating PG$1159$ stars
\citep[e.g.][]{dreizlerheber98, werner01}.

Four model-star sequences, with $0.53$, $0.55$, $0.59$, and $0.64
\msol$, were analyzed in this paper. The $0.59 \msol$ sequence 
derived directly from the evolution computations of
\citet{althausetalvltp05}. Figure~\ref{fig:hrdm27} displays the
evolutionary track of the model with $2.7 \msol$ on the zero-age main
sequence and $0.59 \msol$ at the end on the white-dwarf cooling track.
The stellar models with $0.53$, $0.55$, and $0.64 \msol$ were derived
from the $0.59 \msol$ sequence by changing the stellar mass to the
appropriate value shortly after the termination of the born-again
red-giant phase of the VLTP (i.e. at $\log \teff \approx 4.2$). Upon
restarting the evolution of the models with the abruptly changed mass,
the efficiency of the helium-burning shell did not match the model
structure. For all three stellar masses, the transition phase during
which nuclear burning and the internal structure re-adapted lasted to
about $\log \teff = 5.1$ on the locus of the last descent from the
red-giant domain.

Furthermore, the evolutionary phase along the white-dwarf cooling
track (when $\log\teff < 5.0$) included elemental diffusion which is
important for the occurrence of a red edge of the pulsations as
discussed in Sect.~\ref{sect:rededge}.

%---------------------------------------------------------------------------
%  side-captioned plot
%---------------------------------------------------------------------------

The nonradial stability analyses of dipole and quadrupole \g~modes
were performed with the Riccati integration method of the linear
nonadiabatic oscillation equations as used in the PG$1159$
study of \citet{GaPG97}; the approach was described for example in
\citet{GLFDAV96}. Again, we retained the nuclear terms in the 
stability equations to detect potentially existent $\epsilon$-driven
instabilities. Convective energy transport is not important in the
envelopes of the PG$1159$ stars. Hence, the stability computations
neglected convection effects completely.  The Riccati method was
described on various occasions in the past; nevertheless, elementary
misjudgments continue to prevail \citep[e.g.][]{anc03} and it seems
necessary to emphasize once more that the Riccati approach is a
shooting method. The spatial resolution that can be achieved in the
pulsation computations is not limited by the number of spatial
gridpoints approximating the stellar background model. All that is
required is that the stellar model must resolve all the relevant
physical quantities sufficiently well. The physical quantities that
must be accessible to the pulsation integration at arbitrary spatial
coordinates are then interpolated in the stellar background via a
monotonized cubic polynomial \citep{steffen90}. The limitation of the
radial mode order that was computed in this project (and any other
before) was inflicted by the computing time that had to be invested
per mode rather than by any numerical spatial resolution issue.

In this project, the Brunt-V\"ais\"al\"a (BV) data had to be spatially
smoothed. Very crisp com\-po\-sition steps in the stellar interior, be
they imposed by elaborate treatments of convective boundaries or
artificially produced by the finite-difference methodology, cause
spikes in the BV frequency. Shooting methods are particularly
sensitive to spikes in the coefficients of the differential equations
as they usually entail very short integration steps to spatially
resolve such rapid changes. Arguing that such discontinuities are
somewhat smeared out in a real star, we smoothed the BV frequency
numerically to iron out the spikes but maintaining at the same time
its broader features. The simple algorithm that was invoked computed
the BV frequency iteratively at any point in the star as the average
over its nearest neighbors and consequently removed narrow spikes
efficiently.

%-----------------------------------
\section{Results}
\label{sect:results}
%-----------------------------------
%---------------------------------------------------------------------------
%  full-widthplot
\begin{figure*}
\centering
 \includegraphics[width=18.2cm]{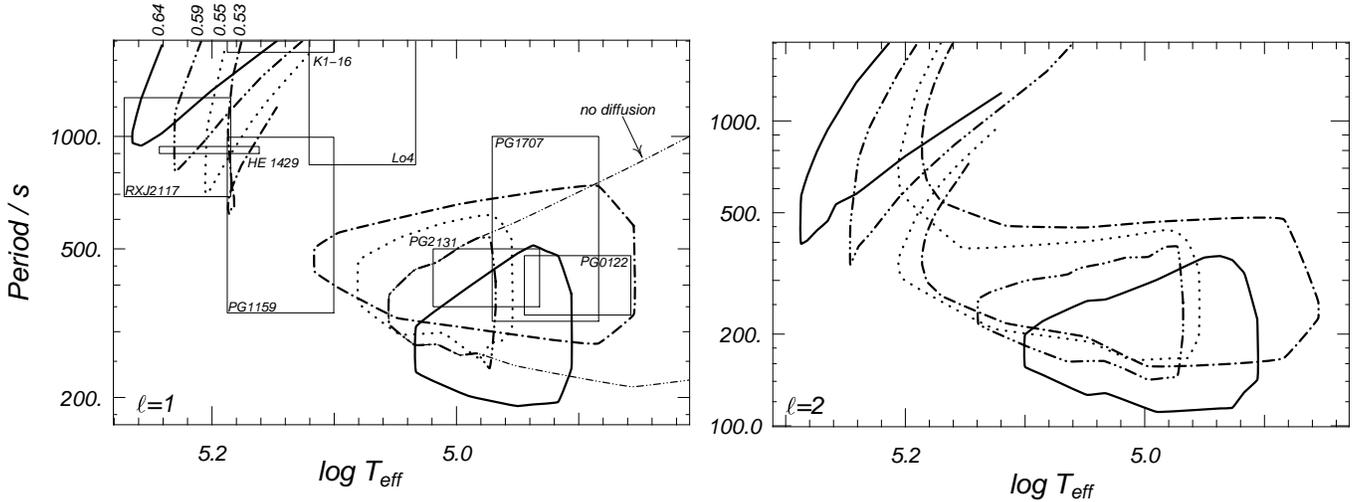} 
 \caption{The
      boundaries of the instability domains on the $\log
      \teff$~--~period domain. The left panel shows the dipole modes,
      the right panel contains the results for the quadrupole modes.
      The loci traced out by the four model sequences are plotted
      with different line types.  For the $0.59\msol$ sequence, the
      results obtained on models with suppressed diffusion are shown
      in addition with a thin dot-dot-dashed line in the left
      panel. The boxes delineate the observational situation for
      dipole modes. Details are given in the text.}
\label{fig:l1l2bounds}
\end{figure*}
%---------------------------------------------------------------------------
This section describes the numerical results obtained from the
nonadiabatic eigenanalyses of $\ell = 1$ and $2$ modes, performed on
the four model-star sequences with $0.53, 0.55, 0.59$, and $0.64
\msol$.

Figure~\ref{fig:l1l2bounds} displays the instability boundaries on the
$\log \teff$~--~period plane for $\ell = 1$ (left panel) and $\ell =
2$ modes (right panel).  Eigenmodes with periods up to $1800$~s were
considered.  For the dipole modes, the observed domains which are
occupied by GW~Vir variables are overlaid for comparison. The observed
dipole-mode period ranges were adopted from Tab.~1 of
\citet{nagelwerner04}; the horizontal sizes of the boxes reflect the
canonical 10~\% uncertainty in the spectroscopic $\teff$
determinations.

The $\log \teff$~--~period diagram for dipole modes shows that the
instability regions are separated into a low-$\log g$ (high
luminosity) and a high-$\log g$ (low luminosity) domain for each
mass. For all stellar masses considered, the tip of the evolutionary
knee is pulsationally stable. The low-$\log g$ instability domains are
roughly compatible with observed periods in variable planetary-nebulae
nuclei (PNNV). One exception is the star Lo4 with its short
periods. The Lo4 observations could possibly be reconciled by adopting
$\mast < 0.53 \msol$.

No red edge was encountered in the low-$\log g$ instability domains.
Due to our restriction of the computations on the long-period side, we
could not follow overstable modes to temperatures below $125\,000$~K
on the high-luminosity branch of the evolutionary tracks.  At these
longest periods that were computed, we did see no signs yet of a
decline of the strength of the instabilities.

The high-$\log g$ instability domains start at $\log \teff = 5.12$ for
the $0.53 \msol$ model and at successively lower $\teff$s for higher
masses. If elemental diffusion was included, all instability domains
showed a sharply defined red edge on the $\log\teff$~--~period plane.
The $0.59 \msol$ model sequence was evolved once with diffusion at
$\log \teff < 5.0$ and once without diffusion. The resulting
instability boundaries are displayed as dot-dot-dashed lines in the
left panel of Fig.~\ref{fig:l1l2bounds}. For $\log\teff > 4.98$ both
instability domains are indistinguishable. The diffusive models
encounter a red edge at $\log\teff \approx 4.97$. The instability
domain of the diffusion-free sequence (thin dot-dot-dashed line), on
the other hand, continues to grow towards lower $\teff$s without a sign
of a weakening at $\log\teff = 4.70$, the location of the coolest
model analyzed.

The position of the red edge is not realistic; its location depends on
the epoch when diffusion is switched on during the evolution
computations. All that is important here is that
\emph{diffusion causes a red edge}; this point is further elaborated
on in Sect.~\ref{sect:rededge}.

The maximal period-widths of the high-$\log g$ instability domains
range from $290$ to $480$~s with no clear mass dependence. On the
other hand, the magnitude of the shortest and the longest overstable
periods are a function of mass.  The $0.53 \msol$ models have the
longest-period overstable modes; the magnitude of the upper period
boundary decrease with increasing mass. For the $0.64 \msol$ sequence,
the longest overstable period is as short as $500$~s.  On the other
side, the shortest-period overstable modes have periods as low as
$192$~s for the $0.64 \msol$ sequence. This is shorter than the
shortest observed periods in any GW Vir star.
%%The excitation rates are, however, at
%%least three orders of magnitude smaller for the short-period modes
%%than for the modes close to the long-period end of the overstable
%%period range.  
 
The most obvious problem is evidently the prototype of the class,
PG$1159$-035 itself. PG$1159$-035 lies so close to the evolutionary knee
of the model tracks that it falls mostly \emph{between} the low- and
high-$\log g$ $\ell = 1$ instability domains. The observational box on
the $\log \teff$~--~period plane overlaps only marginally with the
instability domains of the $0.53 \msol$ sequences. Such a low mass
must be rejected, however, based on the computed period separations
(see also Sect.~\ref{sect:discuss}).

The right-hand panel of Fig.~\ref{fig:l1l2bounds} shows the boundaries
of the instability domains traced out by the quadrupole \g~modes. For
the two most massive sequences, the instability topology looks the
same as for the dipole modes. The main difference is that considerably
shorter-period modes are overstable. For the $0.55$ and $0.53 \msol$
sequences, the instability domains are uninterrupted, they extend
around the knee from the low- to the high-$\log g$ domain. The
effective temperatures of the red edges of the dipole and quadrupole
instability domains coincide.  The growth rates of the overstable
$\ell = 2$ modes reach comparable magnitudes as those of the $\ell =
1$ modes.

%---------------------------------------------------------------------------
%  side-captioned plot
\begin{figure}
      \resizebox{\hsize}{!}{\includegraphics{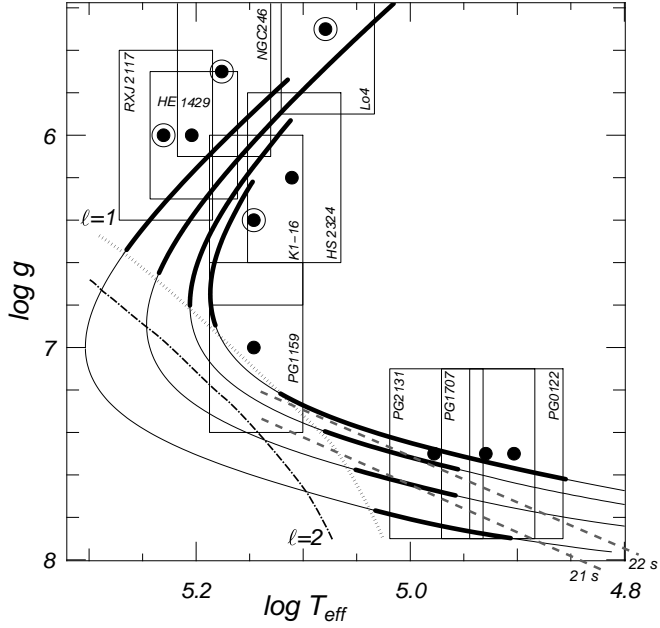}}
      \caption{ The GW Vir instability domains on the $\log \teff -
      \log g$ plane. The dots represent observed variable PG$1159$
      stars. Encircled dots indicate the pulsators that are
      embedded in a planetary nebula. The boxes around the
      observational data mark the canonical uncertainties. The dotted
      line indicates the dipole-mode and the dash-dotted line the
      quadrupole blue edge. Along the evolutionary tracks, the epochs
      with overstable \g~modes are plotted with heavy lines. In the
      lower right of the figure, the two grey dashed lines mark the
      loci of constant period separation ($21$~s and $22$~s) as
      derived from the nonadiabatic stability computations.}
\label{fig:lggtebounds}
\end{figure}
%---------------------------------------------------------------------------
For comparison with spectroscopically calibrated PG$1159$ stars,
Fig.~\ref{fig:lggtebounds} plots the evolutionary tracks and the
overstable regions on the $\log \teff$~--~$\log g$ plane. Observed GW
Vir stars and the associated error boxes are superimposed. PNNVs are
indicated by encircled dots. The loci of the blue edges of $\ell=1$
and $2$ modes are given as dotted and dash-dotted lines,
respectively. The evolutionary phases harboring overstable dipole
\g~modes are marked with a heavier line on top of the evolutionary
track. The general agreement between observations and models is
reasonably good. Note again, the red edge position is not
trustworthy. Important is, however, the existence of a red edge.

%---------------------------------------------------------------------------
%  side-captioned plot
\begin{figure}
      \resizebox{\hsize}{!}{\includegraphics{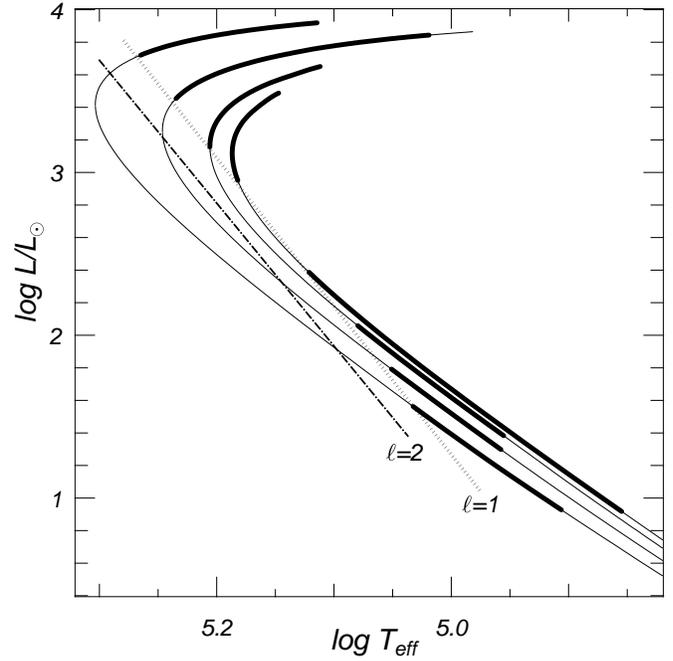}}
      \caption{The loci of all model sequences in the PG$1159$ region of
      the HR plane. The heavy lines delineate the domains of unstable
      $\ell = 1$ \g~modes. The blue edges for dipole and quadrupole
      modes, respectively, are indicated with broken lines. For $\log
      \teff > 5.1$ at high luminosities of the $0.59 \msol$ models, no
      instability is indicated; most probably, however, overstable
      \g-modes exist there. However, the computations were stopped
      because the shortest unstable periods exceeded $1800$~s.}
\label{fig:hrdinsts}
\end{figure}
%---------------------------------------------------------------------------
Finally, Fig.~\ref{fig:hrdinsts} shows the evolutionary tracks of the
model stars on the HR plane; thereon, the blue edges can be
approximated by straight lines.  The dipole-mode blue edge obeys the
relation
\begin{equation} 
\label{eq:dipbluedge}
\log L/\lsol = 9.289 \cdot \log\teff - 45.186\,.    
\end{equation}   
The dash-dotted line in Fig.~\ref{fig:hrdinsts}, delineating the
quadru\-pole-mode blue edge, is less steep than the $\ell = 1$ blue
edge and follows the relation
\begin{equation} 
\log L/\lsol = 9.0372 \cdot \log\teff - 44.158\,.    
\end{equation}   
The above parameterizations of the blue edges imply a \emph{mass
dependence} of the GW Vir instability domains as it is already evident
in Fig.~\ref{fig:l1l2bounds}. In Sect.~\ref{sect:discuss}, we return
to the observation that the orientation of the blue edge on the HR
plane differs from that of the classical instability strip, in
particular the slope as the opposite sign.

%---------------------
\subsection{The excitation mechanism}
\label{sect:excitation}
%---------------------
The \g~modes that are found overstable in our models are all driven by
the $\kappa$-mechanism induced by the opacity bump peaking around
$\log T = 6.2$. Partial ionization of K-shell electrons in C and
O cause the peak. As the matter of the stellar envelopes also contains
heavy elements before the onset of diffusion, the high-temperature
Z-bump enhances the opacity peak around $\log T = 6$.  Hence, the
pulsation driving is the same as in all hitherto stability studies of
PG$1159$ variables.

The abundance profile shown in Fig.~\ref{fig:chipg1159} is
representative of all models before the occurrence of
diffusion-induced helium gradients in the outer envelopes (see
Fig.~\ref{fig:rededge}). For the diffusion-less models, the abundances
remain on the level as at the left border in Fig.~\ref{fig:chipg1159}
out to the photosphere. Note that in Fig.~\ref{fig:chipg1159}, the
temperature reached already $1.6\cdot 10^{7}$~K at $\log(1-m/\mast) =
-5$.  Hence, all model stars have at least $30.6 \%$ of helium (in
mass units) in the driving region. In particular, all oscillating
PG$1159$ models with $\log\teff > 5.0$ (on the white-dwarf cooling
tracks) have \emph{no} composition gradient between photosphere and
the pulsations' driving region. Therefore, the excitation physics
remains the same as discussed by \citet{SaioPG96},
\citet{GaPG97}, and more recently by \citet{quirionetal04}.

%---------------------
\subsection{The red edge}
\label{sect:rededge}
%---------------------
%---------------------------------------------------------------------------
%  side-captioned plot
\begin{figure}
      \resizebox{\hsize}{!}{\includegraphics{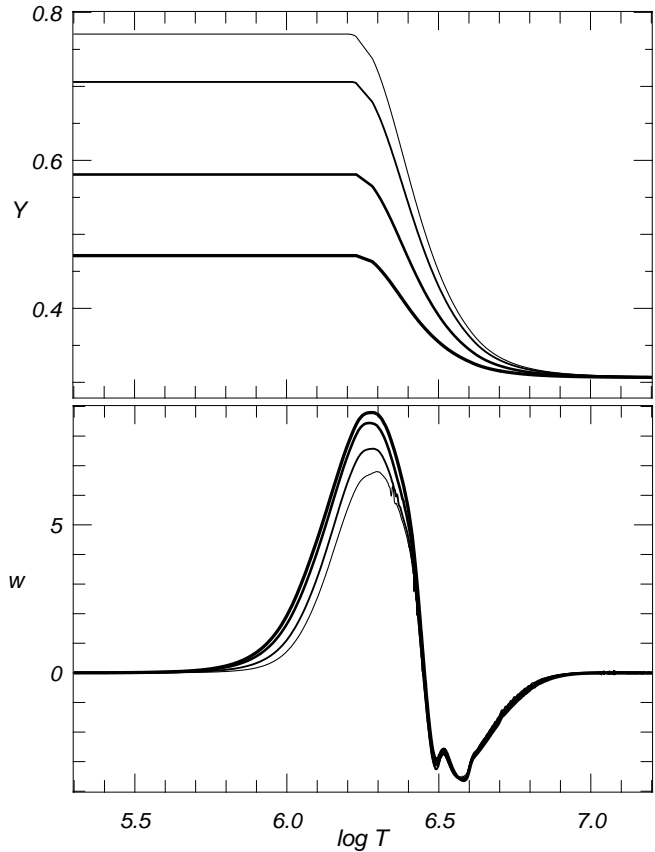}}
      \caption{Top panel: Spatial variation of the He profile (Y) as a
      function of depth, measured in $\log T$, for four $0.55 \msol$
      model stars as the red edge of the \g-mode instability is
      approached. Bottom panel: Differential work, $w \equiv \diff W /
      \diff r$, of the $\ell = 1, k = 19$ mode for the same
      models. The eigenmode belonging to the thinnest line is
      pulsationally stable.}
\label{fig:rededge}
\end{figure}
%---------------------------------------------------------------------------
As mentioned before, all model sequences were evolved with switching
on elemental diffusion after they passed $\log \teff = 5.0$ on the
white-dwarf cooling track. The choice of the epoch of the switch-on of
diffusion was arbitrary. If diffusion alone, without competing
counteracting physical transport processes, is considered,
sedimentation is very effective. The top panel of
Fig.~\ref{fig:rededge} shows the helium profiles close to the \g-mode
driving region at several epochs of the $0.55 \msol$ model sequence.
It took the star to evolve only from $\log \teff = 5.0$ to $4.951$ for
the superficial He abundance to rise from $0.306$ to $0.77$. The
superficial flattening of the He profile is an artifact of the
modeling procedure: the diffusion equations are solved only in the
Henyey part of the stellar interior but not in the stellar envelopes
that are computed in full equilibrium. Hence our results probably
underestimate the diffusion effects.

As mentioned in Sect.~\ref{sect:results}, we computed also the
stability properties of $0.59 \msol$ model stars devoid of
diffusion. The corresponding stability analyses were hence performed
on models with a constant He mass abundance of $0.306$ from the
surface to the nuclear-active deep interior. Neglecting diffusion
altogther for the whole evolutionary history, no damping of the
\g-mode pulsations was found along the white-dwarf cooling track.

On the other hand, if diffusion was accounted for, the \g~mode
instabilities literally starved when the He abundance exceeded a
critical level. The microphysics of elemental transport is such that
the He that is going to float on top of the heavier elements starts to
separate from these other elements at a temperature depth of about
$10^7$~K. The steepest abundance gradients are found between $\log T =
6.7$ and $6.3$, this region also dominates the damping and driving of
the \g-modes for the PG$1159$ variables. Hence, it is understandable
that superseding the main driving elements of that region with
``pulsationally inert'' stellar material reduces the efficiency of the
pulsational driving. Looking at the differential work curves in the
lower panel of Fig.~\ref{fig:rededge} shows indeed that the curves are
essentially self-similar. The higher the helium abundance in the
driving region the less efficient is the driving contribution to the
pulsations. Finally, with Y$= 0.77$ in the superficial layers in
Fig.~\ref{fig:rededge}, the driving cannot compensate the damping
anymore and the \g~modes turn stable. The critical level of helium
above the driving region for the red edge varies between $0.77$ and
about $0.82$ for the different masses of the model stars considered
here.

The differential-work curves are very similar in form as the He
abundance rises towards the surface; the compositional gradient does
not distort the eigenfunctions (the spatial kinetic energy
distribution is essentially unaffected by the diffusion process).
Therefore, one can say that diffusion lets the pulsations ``starve''.
In the present computations, diffusion destroys the pulsations rather
quickly after sedimentation is ``switched on'' at $\teff =
100\,000$~K.  The red edges of the different mass sequences were then
encountered between $70\,800$ and $89\,100$~K. Approaching the red
edge, the long-period modes died away earlier than the short-period
ones. Therefore, close to the red edge, the mean period of the range
of excited modes 

%-----------------------------------
\section{Discussion}
\label{sect:discuss}
%-----------------------------------
%%
%% Topology of the instability domains: Periods, Teffs, log g
%%                 Comparison with observations, Saio and Gautschy
%%
The splitting of the PG$1159$ pulsation domains (see Fig.~
\ref{fig:l1l2bounds}) into two regions~--~a long-period, low-$\log g$
domain and a short-period, high-$\log g$ part~--~agrees with the
results of \citet{SaioPG96} and \citet{GaPG97}. In the current
computations, no continuous instability domain connecting low- and
high-$\log g$ regions was found for dipole modes. In the two older
studies just mentioned, the $0.56$ and $0.57 \msol$ sequences had
continuous instability domains, however. The reason is connected with
the different loci of the evolutionary tracks on the $\teff$~--~$\log
g$ plane, the present models reach higher $\log g$ values earlier in
their evolution than in the old simplified ones of Saio and Gautschy.

Regarding the comparison with observations, there is essentially one
problem case: PG$1159-0354$ itself. As mentioned before, on the
$\teff$~--~period plane, the observational box overlaps only slightly
with the computed instability domains of the $0.53 \msol$ model
sequence.  Comparing the associated computed mean period-separation
($\overline{\Delta \Pi_{\ell=1}} = 22.46$~s) with the observed one
($21.62 $~s) makes it clear that the $0.53 \msol$ model is no
acceptable solution. With regard to the period-separation, the best
agreement was found with the $0.55 \msol$ model having
$\overline{\Delta \Pi_{\ell=1}} = 21.50$~s at $\teff = 138\,000$~K.
This, however, means that the $\ell = 1$ blue edge is at least
$6000$~K to cool there to be compatible with the observed situation.
Our preferred mass for PG$1159-0354$ is closer to the
spectroscopic calibration of \citet{dreizlerheber98} ($0.54 \msol$)
than to the $0.59 \msol$ of \citet{kawalerbradley94}. The latter value
is frequently regarded as kind of a canonical mass for PG$1159-0354$; it
should, though, be kept in mind that the asteroseismic calibrations
rely on fits to some evolutionary models with all their shortcomings
and uncertainties.

According to the current models, regions of exclusively $\ell = 2$
overstable modes were encountered for lower-mass (i.e. about $0.53 <
\mast < 0.57 \msol$) PWDs when they pass along the low-luminosity side
of the evolutionary knee.

As mentioned hitherto for DBVs \citep{alfalt02}, also for the
pulsating PG$1159$ models we found the shortest-period overstable modes
to be quadrupole ones and the longest-period overstable modes to be
dipole ones. Even if the latest RXJ$2117$ observations
\citep{vauclairwetrxj2117} do not clearly reveal quadrupole modes,
the most likely $\ell=2$ candidates in the data are at the
short-period boundary of the observed set. Also for PG$1159$
\citep{wingetwetpg1159}, the positively identified $\ell=2$ mode 
has a short period at $425$~s. However, one multiplet, the one
associated with longest-period mode, is possibly due to an $\ell=2$
mode; if true, this would clearly contradict our computations. Another
potential problem case is PG$0122$; its $465$~s mode~--~the
third-longest period mode reported in \citet{obrienetal96}~--~is
suggested to be a quadrupole mode. The identification was not without
debate, however.

%%
%% Excitation Mechanism: Comparison with previous work
%%
With regard to mode excitation, we saw that neither the use of overly
simplified full stellar models nor the use of envelope models only are
the reason for overstable \g~modes in envelopes with helium mass
abundances of about 0.3. Additionally, we have now three different
numerical methods (finite-difference relaxation, finite-element
solver, and a nonlinear shooting method) that agree on the \g-mode
excitation. The disagreeing results
\citep[e.g.][]{starrfieldetal83,bradleydziem96,anc03}
were obtained with finite-difference relaxation methods which resemble
the method of \citet{SaioPG96}. We hypothesize therefore that the
origin of the disagreement is linked to particularities of the
computational realizations. We found the previously proclaimed
``helium poisoning'' to become effective only at higher mass
abundances such as they are encountered when elemental diffusion was
at work. If the relative mass abundance of helium exceeded about
$0.77$, the \g~modes were found to ``starve'' so that a red edge
occurred.

With respect to pulsational driving, the only qualitative difference
to the results of \citet{SaioPG96} and \citet{GaPG97} is the absence
in the current model stars of $\epsilon$-driven instabilities with
periods of around $100$~s.

%%
%%     Explanation of how to understand the excitation of different
%%     period domains at the same Teff but different L.
%%
\medskip
How can the bimodal period distribution on the $\teff$~--~period plane
(cf. Fig.~\ref{fig:l1l2bounds}) be understood? At a chosen $\teff$,
the periods differ by at least a factor of three, whereas the
luminosities differ by factor of $15$ or more. In contrast to the
expectations, the short periods are excited at low and long periods at
high luminosities.  To understand this, we resort to the classical
time-scale argument
\citep[e.g. in][]{pss}
\begin{equation}
\label{eq:thermdynscale}
\int_{\Delta M}\!\!\! \cv T\,\diff m / \last = {\cal O}(\Pi / 2 \pi)\,.
\end{equation}
The integral extends over the mass of the envelope, $\Delta M$,
overlying driving region of the pulsations. Relation
(\ref{eq:thermdynscale}) defines the pulsation period, $\Pi$, for
which the thermo-mechanical coupling in the stellar material is
optimal for pulsational driving. We compared two $0.59 \msol$ models
at $\log \teff = 5.06$, one at high and the other at low
luminosity. For both models, we evaluate the left-hand side in
eq.~(\ref{eq:thermdynscale}). We approximate the integral, as usual,
by adopting the values at the driving region for the quantities $\cv$
and $T$ (indicated by a tilde) and write it hence as
$\tilde{\cv}\,\tilde{T}\,\Delta M$.  Based on the full stability
computations, we assumed the driving to occur at around $\log
\tilde{T} = 6.3$ in all cases. Therefore, we compute the 
ratios of $\tilde{\cv}, \Delta M$, and $\last$; all three quantities
are considerably larger than unity as the evolution proceeds past the
knee (always measured at the same $\teff$). It turns out that the
product in the numerator of Eq.~(\ref{eq:thermdynscale})
over-compensates the change in $\last$ in the denominator. Therefore,
the high-luminosity phase of evolution prefers longer periods for
pulsation than the low-luminosity one. The overcompensation is mainly
due to the considerable decrease of the envelope mass, overlying the
driving region, as the stars evolve towards to white-dwarf cooling
track. At $\log \teff = 5.06$ for example, $\Delta M$ at high $\last$
is 630 times larger than at low $\last$. The influence of this number
is further enhanced by the fact that radiation pressure constitutes a
considerable fraction of the total pressure, $P$, in the
high-luminosity stage. Since $\cv \propto 1/\beta^2$, with $\beta = 1
- \Prad/P$, $\cv$ can rise considerably when radiation pressure gains
importance. Again, for the models at $\log \teff = 5.06$ we found
$\tilde{\beta} \approx 0.4$ at the high luminosity and $\tilde{\beta}
\approx 0.9$ at the low-luminosity epoch.

\medskip

To understand the slope/orientation of the blue edge in the post-knee
domain in the HR diagram, we follow the reasoning of
\citet{coxhansen79}.  We approximate eq.~(\ref{eq:thermdynscale}) the
same way as to understand the period bimodality. We approximate the
hydrostatic equilibrium as

\begin{equation*}
\frac{4 \pi \tilde{P} R^4}{G M} \approx \Delta M \,.
\end{equation*}
The Stefan-Boltzmann relation allows to rewrite
eq.~(\ref{eq:thermdynscale}) as
\begin{equation*}
 \frac{\tilde{\cv} \tilde{T} \tilde{P} L}{\Pi\,\teff^8\,M} 
      \approx \text{const.}
\end{equation*}
The envelope $P - T$ structure of PWDs is decently approximated by a
radiative-zero one using a Kramers-type opacity.  Eliminate
$\tilde{P}$ in the last equation with the chosen $P - T$ relation to
get
\begin{equation}
\label{eq:thermech}
 \frac{\tilde{\cv} \tilde{T}^{5.25}}{\Pi\,\teff^8}\cdot
 \sqrt{\frac{L}{M}} \approx \text{const.}
\end{equation}
For the period $\Pi$, we adopt the relation $\Pi \propto
T_m^{-\frac{1}{2}} R$ from \citet{wingethnsnvnhrn83}.  Using a
very simple white-dwarf model from \citet{kw} (their section 35.3):
$T_m \propto \left({L}/{M}\right)^{{2}/{7}}$ to
estimate the maximum temperature, $T_m$, leads to
\begin{equation*}
\Pi \propto R\,\left( \frac{L}{M} \right)^{-\frac{1}{7}}\,.
\end{equation*}
Eliminating $R$ again via Stefan-Boltzmann's law gives
\begin{equation}
\label{eq:thermech1}
 \frac{\tilde{\cv} \tilde{T}^{5.25}}{\teff^6}\cdot
 \frac{L^\frac{1}{7}}{M^\frac{9}{14}} \approx \text{const.}
\end{equation}
Assume the temperature at the top of the driving region, $\tilde{T}$,
to remain constant within the instability domain of the PG$1159$
variables and neglect the mass dependence of the derived relation, as
the range of stellar masses assigned to PG$1159$ variables is small.
From fits to the evolutionary computations we derive
$\diff\ln\tilde{\cv} / \diff\ln L\vert_{\teff}\approx 0.5 $ and we
neglect any additional $\teff$ dependence. All things considered, we
finally arrive at:
\begin{equation}
\frac{\diff\ln L }{\diff\ln \teff} = \frac{28}{3}\approx 9.3 \,.
\end{equation}
According to eq.~(\ref{eq:dipbluedge}), the slope of the blue edge as
obtained from the fit to the full nonadiabatic computations is $9.3$
for $\ell = 1$ and $9.0$ for $\ell = 2$. Regarding the simplicity of
the model and the coarseness of the approximations, the result is
astonishing. We attribute the dominant contribution to the slope of
the blue edge to the term $\diff\ln \cv / \diff\ln L$. Hence, it is
the growing importance of radiation pressure as the luminosity
increases in the PG$1159$ envelopes that tilts the blue edge to the left
on the HR- and the $\log \teff - \log g$ diagrams.

%%
%% Red edge: Position now, with ad hoc switch-on of diffusion. 
%%           Self consistent treatment needed in the future.
%%           It is conceivable that observations serve as the
%%           constraints on the efficiency of different counteracting
%%           mechanisms that finally define red edge on a star to star
%%           basis rather than one red edge for the class as a whole.
%%

We found elemental diffusion to quench g-mode instability on the
high-$\log g$ branch of evolution when the helium abundance exceeds a
critical level in the driving region. A helium mass abundance of about
$0.77$ was found to be sufficient for the overstable \g~modes to
starve. Hence, as conjectured by \citet{dreizlerheber98}, a depletion
of C and O, which can be regarded complementarily as an enhancement of
He, due to sedimentation in the driving region enforces a red edge.

Diffusion is a very efficient process to turn PG$1159$ stars into DO
stars. If only diffusion~--~without counteracting processes~--~is
included in stellar modeling, DO stars would show up at too high
effective temperatures to be compatible with observations.  In the
current modeling, diffusion was switched on ad hoc at $100\,000$~K on
the white-dwarf cooling track.  Therefore, the location of the red
edge of the PG$1159$ variables in the various diagrams shown in this
paper is not trustworthy.

The quantitative non-equilibrium computations of \citet{unglaubues00}
to model the counteracting processes of diffusion and mass loss
suggest that a unique red edge for the PG$1159$ variables as a
\emph{class} cannot be expected.  As compositions and rotational
velocities (which tend to homogenize composition gradients) vary from
star to star, also the epoch (and hence the corresponding $\teff, L$,
or $g$) will vary when the critical contamination of the driving
region is reached. Therefore, it is not inconceivable that close to
the red edge of the PG$1159$ variables, these are intermixed with
stable DO stars and even with non-pulsating PG$1159$ stars. The closer
one gets to the red edge, the shorter should the mean period of the
excited modes be; this occurs since, according to our computations,
the long-period modes starve earlier than the short-period ones.

The position of the theoretical instability domains is only part of
the game to understand the GW Vir pulsators. Pertinent information is
obtained from period-spacing data. We used the results from $\ell = 1$
computations to compare them with the observed period separations for
three stars. In the case of RXJ$2117+3412$, only tracks of $0.64$ and
$0.59 \msol$ models pass on the $\log \teff - \log g$ plane within the
observational error box.  Inspecting the period separations revealed
that the 0.64 models have too small a mean period separation. Even the
$0.59 \msol$ objects with the arithmetic-mean period separation
$\overline{\Delta \Pi_{\ell=1}} \approx 20.0$~s were below the mean
observed separation of $21.62$~s
\citep{vauclairwetrxj2117}. Hence, only a mass below $0.59 \msol$ can
fit the observation. However, the corresponding current evolutionary
tracks pass outside the error box.  PG$1159-035$'s observed
$\overline{\Delta \Pi_{\ell=1}} = 21.62$~s period spacing
\citep{wingetwetpg1159} is best reproduced with our $0.55 \msol$
models.  As mentioned before, the problem with the $0.55 \msol$ model
stars is that none of the involved dipole modes is pulsationally
overstable in the vicinity of the position of PG$1159-035$ on the
$\log\teff - \log g$ plane. Again, evolutionary models that arrive at
the lower side of the knee with higher luminosity than what we get
currently might resolve the dilemma. Finally in PG$2131+066$,
$\overline{\Delta \Pi_{\ell=1}} = 21.65$~s is observed
\citep{kawetal95}. From our current modeling, we conclude that the
best fitting stellar mass lies between $0.55$ and $0.59 \msol$; the
observed pulsation modes fall then well within the computed range of
overstable dipole \g~modes (cf. Fig.~\ref{fig:l1l2bounds}).

Figure~\ref{fig:lggtebounds} has inscribed lines of constant period
separations. The loci of $21$~s and $22$~s, interpolated from the
nonadiabatic computations, are shown. Note that the observed period
separations for RXJ$2117$, PG$1159-035$, PG$2131$, and
\citep[according to][]{obrien00} PG$0122$ all lie between 
$21$ and $22$~s. If PG$1707+427$'s period separation of $18.37$~s 
\citep{fontaineetal91} holds then the apparent confinement to 
$21 - 22$~s of the other four GW Vir variables is an observational
bias.  However, in case the period separations of GW Vir stars are
indeed narrowly confined, then they populate only a fraction of the
instability region as computed by linear theory. It appears worthwhile
to point out that the loci of constant period separation depend also
on the trapping amplitude of the modes. Our mean period separations
were computed as simple arithmetic means over many successive radial
orders. The larger the trapping amplitudes are, i.e. the sharper the
composition gradients in the stellar interior, the bigger the
difference between $\overline{\Delta \Pi}$ and the asymptotic
value. If mass-dependent diffusion efficiencies are capable to
increase the distance between the, say $21$ and $22$~s,
period-separation lines in Fig.~\ref{fig:lggtebounds} remains to be
seen. On the observational side, the number of as\-tero\-seismically
studied PG$1159$ is so low, that every additional well-resolved
oscillation-frequency spectrum that becomes available makes an
important contribution to solve this conundrum.

In addition to the period-separation information, we can also compute
the lengths of the trapping cycles. The two comparisons with
observations that are currently possible revealed the modeled
trapping-cycle lengths to be always shorter than the observed ones:
RXJ$2117+3412$, $83.9$~s observed and $56 - 60$~s computed. For
PG$1159-035$: $80.5$~s observed and $52.6 - 68.4$~s computed. Based on
mode-trapping theory as developed in \citet{brassetal92}, the current
model stars have seemingly too thick helium layers to be compatible
with observations. As there is ample evidence for mass loss before and
during the PG$1159$ episodes of evolution \citep{werner01}, we are
positive that the mass of the helium-rich model envelopes can be
sufficiently reduced via a hot stellar wind to improve the agreement
with observations.

%-----------------------------------
\section{Conclusions}
\label{sect:conclude}
%-----------------------------------
The excitation physics as scraped out from simplified model stars by
\citet{SaioPG96} and \citet{GaPG97} is corroborated using now complex
stellar-evolution models that went through a very late thermal flash.
In particular, we found again that modes can be excited with
considerable He abundance in the driving region, supporting the view
that abundance gradients between the spectroscopically accessible
stellar surface and the driving region of \g~modes do not have to be
invoked to explain PG$1159$-type oscillations
\citep[e.g.][]{bradleydziem96}.

A blue edge was encountered for dipole and quadrupole modes; its slope
is a manifestation of a mass dependence of the pulsational instability
of PG$1159$ stars. On the high-luminosity branch of the evolutionary
track, the hottest variables are the most massive ones. The situation
reverses on the low-luminosity branch (below the evolutionary knee),
where the bluest variables are the least massive ones.
 
Model-star sequences including diffusion led to a red edge of the
PG$1159$ instability domain. Much in agreement with the predictions of
\citet{dreizlerheber98}, the oscillations of the PG$1159$ stars 
starve once the He abundance in the driving region exceeds a
critical magnitude. It is therefore likely that the red edge is close
to the transition from PG$1159$ to hot DO- or possibly DA-type white
dwarfs. It is furthermore conceivable that the red edge of the GW Vir
stars is not sharply defined but rather frayed.  The counteracting
effects of rotation, stellar wind, radiative levitation and
gravitational settling determine for each star individually when the
\g-mode pulsations die out. A significant step ahead would be
parameter studies of the competition of mass-loss and/or diffusion on
the efficiency of pulsational driving.  Currently, diffusion is still
switched on ad-hoc.

In contrast to the models of \citet{SaioPG96} and \citet{GaPG97}, the
stability analyses of the current stellar model sequences did not
reveal any overstable $\epsilon$-mechanism driven \g~modes at short
periods. 

We gathered some evidence, when comparing the theoretical with the
observed period-separation data, that the current evolutionary tracks
evolve at too low luminosity on the HR diagram and hence on the
$\log\teff - \log g$ plane. Furthermore, based on the magnitudes of
the trapping-cycle lengths, the helium-rich envelopes of the current
PG$1159$ models seem to be too massive to agree with observations.  It
appears therefore worthwhile to look again into the microphysics and
the numerics of evolutionary computations to try to resolve the
persisting discrepancies.

%-----------------------
\begin{acknowledgements}
%-----------------------
%%% A.G. was not financially supported by any governmental
%%%science foundation whatsoever.
L. G. A. acknowledges the Spanish MCYT for a Ram\'on y Cajal
Fellowship.  We are indebted to K. Werner for explaining to us
observational matters of relevance. A.~Corsico helped us with
discussions concerning the spatial smoothness of the
Brunt-V\"ais\"al\"a frequency. This research has made use of NASA's
Astrophysical Data System Abstract Service.

%-----------------------
\end{acknowledgements}
%-----------------------

%%%%%%%%%%%%%%%%%%%%%%%%%%%%%%%%%%%%%%%%%%%%%%%%%%%%%%%%%%%%%%%%%%%%%%%%%%%%%%
% The Bibliography
%-----------------------------
\bibliographystyle{aa}
\bibliography{StarBase}       
%------------------------------------------------------------------------------
\end{document}